\newcommand{\avg}[1]{\left< #1 \right>} 

\documentclass{aastex}
\usepackage{amsbsy}
\usepackage{amsfonts}
\usepackage{amssymb}
\usepackage{longtable}
\usepackage{lscape}

\usepackage{graphicx}
\usepackage{graphics}
\usepackage{dcolumn}
\usepackage{bm}
\begin{document}

\title{Evidence of Wave Damping at Low Heights in a Polar Coronal Hole}
\author{M. Hahn\altaffilmark{1}, E. Landi\altaffilmark{2}, and D. W. Savin\altaffilmark{1}}

\altaffiltext{1}{Columbia Astrophysics Laboratory, Columbia University, MC 5247, 550 West 120th Street, New York, NY 10027 USA}
\altaffiltext{2}{Department of Atmospheric, Oceanic and Space Sciences, University of Michigan, Ann Arbor, MI 48109}

\date{\today}
\begin{abstract}
	We have measured the widths of spectral lines from a polar coronal hole using the Extreme Ultraviolet Imaging Spectrometer onboard \textit{Hinode}. Polar coronal holes are regions of open magnetic field and the source of the fast solar wind. We find that the line widths decrease at relatively low heights. Previous observations have attributed such decreases to systematic effects, but we find that such effects are too small to explain our results. We conclude that the line narrowing is real. The non-thermal line widths are believed to be proportional to the amplitude of Alfv\'en waves propagating along these open field lines. Our results suggest that Alfv\'en waves are damped at unexpectedly low heights in a polar coronal hole. We derive an estimate on the upper limit for the energy dissipated between 1.1~$R_{\sun}$ and 1.3~$R_{\sun}$ and find that it is enough to account for up to 70\% of that required to heat the polar coronal hole and accelerate the solar wind. 
\end{abstract}

\maketitle
	
\section{Introduction} 
	
	Despite more than 50 years of investigation, two main aspects of the solar corona have not yet been fully explained, namely its heating and the resulting acceleration of the solar wind. Two broad classes of models have been proposed for coronal heating and solar wind acceleration: wave driven models, which transfer energy to the corona and solar wind via waves and turbulence; and reconnection driven models in which energy stored in the magnetic fields is released into the corona as the fields relax to lower energy states through magnetic reconnection \citep{Cranmer:LRSP:2009}. 
	
	Wave-driven models benefit from the fact that waves have also been detected in coronal and interplanetary plasma from the chromosphere to 1~AU \citep{Banerjee:SSR:2011}. In particular, Alfv\'en waves appear to be ubiquitous in the Sun. These are transverse MHD waves that travel along the magnetic field lines. They have been detected in chromospheric bright points \citep{Jess:Sci:2009}, spicules \citep{DePontieu:Sci:2007,McIntosh:Nature:2011}, solar prominences \citep{Okamoto:Sci:2007}, the solar corona \citep{Tomczyk:Sci:2007}, and the solar wind \citep{Belcher:JGR:1971}. They have also been found to have low frequencies, with periods on the order of $\sim 100$~s, roughly matching the timescales of the observed photospheric motions from which they presumably originate \citep{Bahng:ApJ:1961,Leighton:ApJ:1962}. And they seem to carry sufficient power to heat the corona and thereby accelerate the solar wind \citep{DePontieu:Sci:2007}. 
	
	Strictly speaking, some of these observations are probably of fast kink waves, not pure torsional Alfv\'en waves \citep{VanDoorsselaere:ApJ:2008}. However, \citet{Goossens:AA:2009} emphasizes that kink waves in the corona have mixed properties and a better label would be ``Alfv\'enic''. For the discussion of waves throughout this paper we adopt the label Alfv\'en to describe the waves, with the caveat that we are using the term broadly. 
	
	Alfv\'en waves are expected to dissipate primarily by collisional processes such as viscosity, thermal conductivity, and resistivity \citep{Parker:ApJ:1991,Cranmer:SSR:2002}. However, these processes have characteristic damping lengths of about 2 to 5~$R_{\sun}$. These waves may provide the sustained energy deposition required by models to match observed solar wind acceleration and proton temperatures. However, predictions indicate that they do not damp below $\approx 2$~$R_{\sun}$ and Coulomb collision rates are too low for any energy dissipated above 2~$R_{\sun}$ to be conducted downward to heat the base of the corona.
	
	Some theoretical work has also shown that Alfv\'en waves can be strongly damped in an inhomogenous plasma through phase mixing \citep{Heyvaerts:AA:1983}, turbulent cascade \citep{Matthaeus:ApJ:1999}, or resonant absorption \citep{Goossens:SSR:2011}. If such dissipation were present, Alfv\'en waves would be a viable mechanism for coronal heating and solar wind acceleration. However, to date there has been no unambiguous observational evidence in the solar corona for wave damping at low heights, which we define here as $<1.4$~$R_{\sun}$; and it is not clear if the conditions required by these models are actually present in the Sun. 

	A common observational approach for detecting the signatures of wave damping is to look at the non-thermal broadening of optically thin spectral lines. The unresolved motions responsible for this broadening are usually attributed to Alfv\'en waves and predicted to be proportional to the wave amplitude \citep[e.g.,][]{Banerjee:AA:1998,Doyle:SolPhys:1998,Moran:AA:2001, Banerjee:AA:2009}.	
	
		Here, we focus only on observations of polar coronal holes, which are mainly open field regions. Coronal holes are the source region of the fast solar wind \citep{Krieger:SolPhys:1973}. One reason for studying these regions is that the fast wind is much less influenced by Coulomb collisions than the slow wind, so that the signatures of other processes at play can be more readily observed \citep{Hollweg:JAA:2008}. 
		
		Many observations have shown that the line widths in coronal holes initially increase with height. This is also predicted for undamped waves based on energy conservation considerations \citep{Hassler:ApJ:1990,Moran:AA:2001}. Some studies, though, have found that the widths level off or even decrease above $\sim 1.1$~$R_{\sun}$, which implies that the waves are damped at heights lower than expected \citep{Banerjee:AA:1998,Doyle:AA:1999,Moran:ApJ:2003,Oshea:AA:2005,Dolla:AA:2008}. Such damping could provide a heating source for coronal plasma near where the fast wind begins to be accelerated. However, the authors of all these studies have argued that systematic effects have rendered their results inconclusive. For example, some of the observations were carried out using the Solar Ultraviolet Measurement of Emitted Radiation Spectrometer \citep[SUMER;][]{Wilhelm:SolPhys:1995} onboard the \textit{Solar and Heliospheric Observatory} (\textit{SOHO}). These were significantly affected by instrumental scattered light (described in Section~\ref{subsec:scat}) and \citet{Moran:ApJ:2003} and \citet{Dolla:AA:2008} attributed the line width decrease to a systematic effect due to scattered light. \citet{Moran:ApJ:2003} noticed that lines from ions with different formation temperatures behaved differently, which suggests a line-of-sight effect in which the observed emission comes from different temperature regions. In another measurement, using the Coronal Diagnostic Spectrometer \citep[CDS;][]{Harrison:SolPhys:1995} also onboard \textit{SOHO}, \citet{Oshea:AA:2005} found a decrease in line widths beginning at about the same height where photoexcitation became significant. They suggested there may be a possible non-damping physical explanation for the line width decrease, but the exact mechanism proposed was unstated. All these potential systematic effects have left unresolved whether the observed line width decrease is real or not.

	Here, we present unambiguous evidence that the decrease is real for the line widths in polar coronal holes. Our measurements address the uncertainties noted in previous studies. We also consider line-of-sight issues and extend the observations to 1.4~$R_{\sun}$, a height larger than that reached with either CDS or SUMER. The rest of this paper is organized as follows: Section~\ref{sec:obs} describes the observations and Section~\ref{sec:ana} describes the analysis methods. Our results are reported in Section~\ref{sec:res} with a discussion of the various possible systematic effects. A possible interpretation of these results in terms of Alfv\'en waves is given in Section~\ref{sec:dis}. Section~\ref{sec:sum} concludes.
		 	 
\section{Observations}	\label{sec:obs}
	
\begin{figure}
\centering \includegraphics[width=0.9\textwidth]{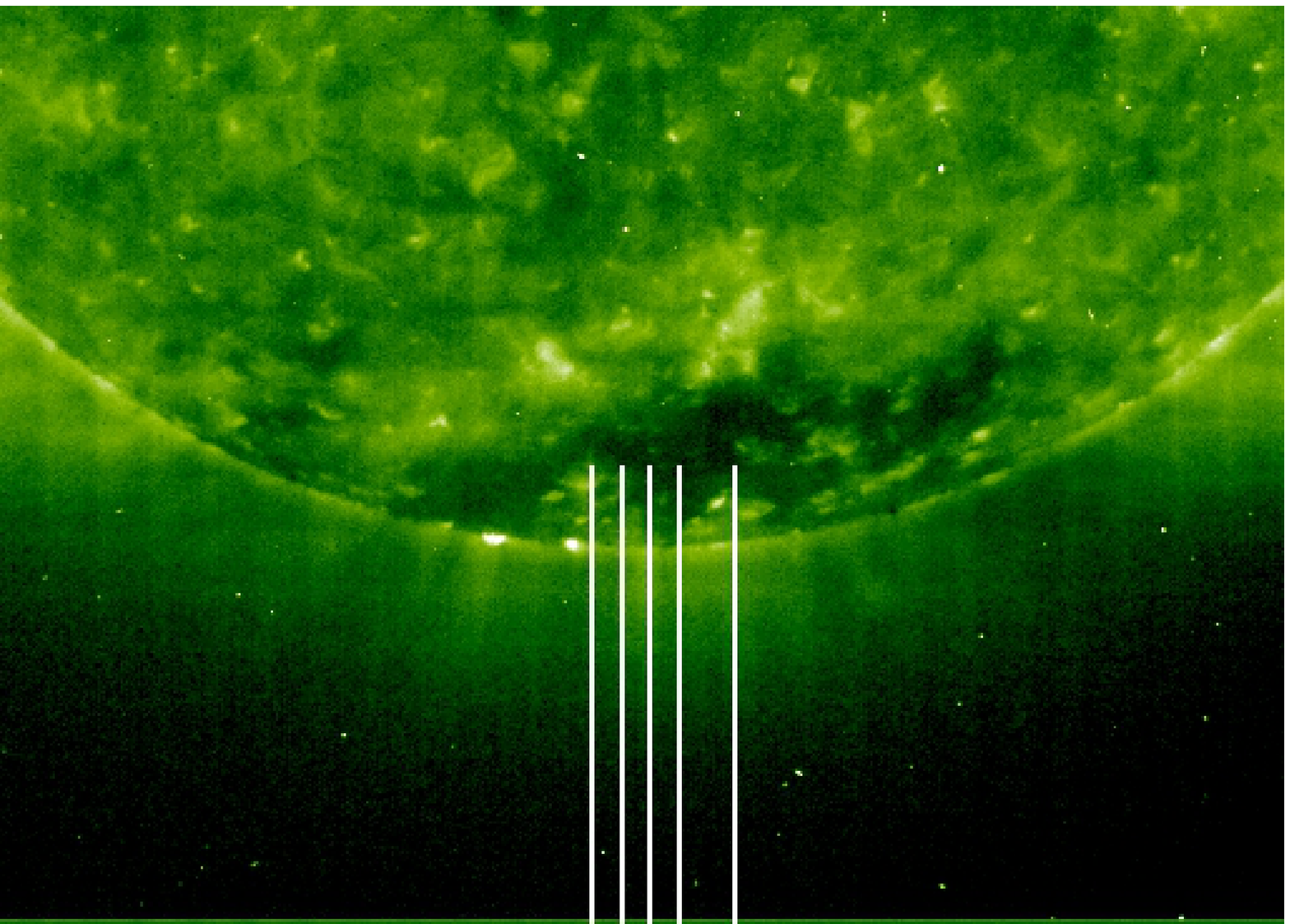}
\caption{\label{fig:context} The slit positions of the EIS observations are illustrated here overlayed on an EIT/\textit{SOHO} image in the 195~\AA\ band, which consists primarily of Fe~\textsc{xii} emission. 
}
\end{figure}

	We have combined data from five observations made with the Extreme Ultraviolet Imaging Spectrometer \citep[EIS;][]{Culhane:SolPhys:2007} onboard \textit{Hinode} \citep{Kosugi:SolPhys:2007}. These observations were made on 2009 April 23 at times 12:08, 12:42, 13:16, 13:50, and 15:17 UT using the $2\arcsec$ slit, which was positioned relative to the central solar meridian at, respectively, $X=-44.5\arcsec$, $-14.5\arcsec$, 15.5$\arcsec$, 45.4$\arcsec$, and 105.6$\arcsec$ (Figure~\ref{fig:context}). In each case the vertical center of the $512\arcsec$ long slit was set at about $-1140\arcsec$ so that the height range included in the observations extended from about 0.95~$R_{\sun}$ to 1.45~$R_{\sun}$. All portions of the observations were within the boundaries of the south polar coronal hole. The exposure integration time for each pointing was about 30 minutes.
	
	Images taken the same day by the Extreme Ultraviolet Imaging Telescope \citep[EIT;][]{Delaboudiniere:SolPhys:1995} onboard \textit{SOHO} were inspected to assess the presence and potential importance of plume plasmas in the EIS field of view. The field of view did not include any significant plume material at any of the slit positions, and so the EIS observations can be considered to consist essentially of only interplume plasma. 
	
	The five observations were averaged together in order to improve the statistical accuracy. To do this, each dataset was first processed using the standard EIS analysis software to remove spikes, warm pixels, and the CCD dark current. Warm pixels flagged by the calibration routine were interpolated using the recommended method, which has been shown to accurately reproduce the line profiles \citep{Young:EISNOTE13}. Systematic drifts in the wavelength scale were then corrected. The separate observations were combined by averaging over pixels located at the same radius. Finally, the data were binned in the vertical direction into bins of 32 pixels each. 

\section{Analysis} \label{sec:ana}
\subsection{Line Fitting} \label{subsec:fit}

	We fit Gaussian profiles to the spectrum in order to derive line widths $\Delta \lambda$ from these data. Fitting is required to resolve the changes in $\Delta\lambda$, which here are on the order of one tenth of the spectral pixel separation of 0.022~\AA$\,\mathrm{pixel^{-1}}$. We tested the accuracy of our fitting procedures by generating synthetic data with a known $\Delta \lambda$, adding to both the line shape and the background a distribution of random noise, corresponding to that seen in the observations, and then running the synthetic line through our fitting procedure. Our analysis showed that, on average, the fitting procedure reproduces $\Delta \lambda$ to better than about $0.1$~m\AA. However the statistical uncertainty of the individual fit parameters was underestimated compared to the standard deviation found by performing the test repeatedly with different random errors all drawn from the same distribution. There could be several reasons for this. One is that the error bars on the intensity data outside the emission line are smaller than the level of background scatter. These intensity data error bars are derived from the EIS preparation routines and the weighted averaging from the binning. If these errors are too small they could cause some pixels to be weighted too strongly in the fit. Another possibility is that the assumed fitting function may not be an exact representation of the data. For example, there could be weak lines in the spectrum, while we assume a constant or linear background level that does not account for them. 
	
	To account for these issues in the data analysis, we derived line width uncertainties from a Monte-Carlo error analysis. After performing an initial fit to the data, normally distributed random numbers were added to each data point. These values were scaled so that the distribution from which they were generated had a standard deviation equal to the residual between each point and the initial fit. Thus, weak features not accounted for by the model fitting function are treated as noise. This accounts for systematic errors when the fitting function is not a perfect representation of the data. The perturbed data were fit and the process was repeated with different random numbers drawn from the same distribution. The uncertainties on the parameters were determined by the standard deviation of the results from many fits. We found that this gave error bars that were similar in magnitude to the scatter seen in the $\Delta\lambda$ results versus radial height $R$. Thus, this procedure produces a more reasonable representation of the actual uncertainties.
	
	An additional possible source of uncertainty is that there might be warm pixels that are not identified as such in the calibration. The warm pixels maps for the EIS calibration are created roughly weekly by identifying those pixels that appear anomalously bright in dark frame exposures. These maps show that the warm pixels are distributed randomly in both spatial and spectral dimensions. It is possible that there are pixels with a warm pixel character that are too weak to appear in the maps. Since any unflagged warm pixels are weak by definition it is unlikely that they have a significant effect on the inferred line profile. To confirm this, we evaluated their effect on inferred line widths by adding random warm pixels to a synthetic spectrum. On average, the spurious pixels tended to slightly broaden the lines, but within the fitting uncertainty. In most cases a randomly placed warm pixel falls far enough away from the line that it has negligible effect on the inferred profile. Because any possible mildly warm pixels fall randomly in space their influence is also mitigated by the 32 pixel spatial binning that we used. Thus, the possible existence of warm pixels not accounted for in the calibration is not expected to systematically alter the inferred line parameters. 

\subsection{Line Widths} \label{subsec:wid}		
	
		The observed width of an optically thin emission line depends on instrumental broadening $\Delta \lambda_{\mathrm{inst}}$, the ion temperature $T_{\mathrm{i}}$, and the non-thermal velocity $v_{\mathrm{nt}}$ \citep{Ultraviolet}:  
\begin{equation}
\Delta \lambda_{\mathrm{FWHM}} = \left[ \Delta\lambda_{\mathrm{inst}}^2 +  
4 \ln(2)\left(\frac{\lambda}{c}\right)^{2}\left(\frac{2k_{\mathrm{B}}T_{\mathrm{i}}}{M} + v_{\mathrm{nt}}^2 \right) \right]^{1/2}. 
\label{eq:width}
\end{equation}
Here $\lambda$ is the line wavelength, $c$ is the speed of light, $k_{\mathrm{B}}$ is the Boltzmann constant, and $M$ is the mass of the ion. The thermal plus non-thermal full width at half maximum (FWHM) is typically 0.04 -- 0.06~\AA, and the instrumental width is about 0.06~\AA. 

The line width includes the sum of the thermal and non-thermal velocities. This 
causes some ambiguity about whether changes in $T_{\mathrm{i}}$ or $v_{\mathrm{nt}}$ are 
responsible for any observed radial variation. However, in a polar coronal hole $T_{\mathrm{i}}$ 
should increase with height due to ion cyclotron resonance heating of the ions by waves with frequencies close to the ion gyrofrequency \citep{Hollweg:JAA:2008}. This heating is balanced in the low corona by 
collisions with cooler protons.	Previous measurements of $T_{\mathrm{i}}$ are 
consistent with the ion temperature being constant or moderately increasing 
at low heights in coronal holes \citep{Landi:ApJ:2009, Hahn:ApJ:2010}. At larger 
heights collisional cooling becomes ineffective and $T_{\mathrm{i}}$ rises 
dramatically \citep{Esser:ApJ:1999}. For these reasons any decreases in line 
width can be attributed solely to decreases in $v_{\mathrm{nt}}$. Thus, for the 
purposes of this work it is sufficient to consider the effective velocity
\begin{equation}
v_{\mathrm{eff}}=\sqrt{\left(\frac{2k_{\mathrm{B}}T_{\mathrm{i}}}{M} + v_{\mathrm{nt}}^2 \right)}. 
\label{eq:veffdefine}
\end{equation}
This is the line width expressed as a velocity after subtracting off the 
instrumental broadening.
	
	The EIS instrumental width is of the same order or even larger than the individual thermal and non-thermal line widths. The pre-launch laboratory measurements of \citet{Korendyke:ApplOpt:2006} showed that the instrumental width for the $1\arcsec$ slit of EIS was $\Delta\lambda_{\mathrm{inst}} = $ 0.047~\AA\ in the short wavelength band (170 -- 210~\AA) and $\Delta\lambda_{\mathrm{inst}} = $ 0.055~\AA\ in the long wavelength band (240 -- 290~\AA). \citet{Brown:ApJ:2008} found that the orbital instrumental width is slightly broader with $\Delta\lambda_{\mathrm{inst}} = $ 0.054~\AA\ and 0.055~\AA\ in the short and long wavelength bands, respectively. A comparison between observations of the same location measured with both the $1\arcsec$ and $2\arcsec$ slits showed that the $2\arcsec$ slit has a $\Delta \lambda_{\mathrm{inst}}$ that is 0.007~\AA\ broader than for the $1\arcsec$ slit \citep{Young:EIS:2011}. That is, the instrumental widths in the short and long wavelength bands for the $2\arcsec$ slit are 0.061~\AA\ and 0.062~\AA, respectively. 
	
	It is also possible that the instrumental width varies along the length of the slit. This has been studied by \citet{Young:EIS:2011} and \citet{Hara:ApJ:2011}. \citet{Young:EIS:2011} measured the line widths above the solar limb in equatorial regions and showed that $\Delta \lambda_{\mathrm{inst}}$ has a U-shaped dependence on location along the slit. However, because the radial distance from the limb varied from $\approx 1.05$~$R_{\sun}$ at the center of the slit to $\approx1.2$~$R_{\sun}$ at the ends of the slit, some of the apparent variation in $\Delta \lambda_{\mathrm{inst}}$ may actually have been from radial variation in $v_{\mathrm{nt}}$. Additionally, the magnitude of $\Delta \lambda_{\mathrm{inst}}$ from this study is larger than expected, being about $0.066$~\AA\ at the center of the $2\arcsec$ slit. \citet{Hara:ApJ:2011} performed a cross calibration of the EIS instrumental width by comparing to ground based observations. They only studied a limited range of the EIS detector, but did find qualitiatively the same U-shaped behavior as \citet{Young:EIS:2011}. 
		
	For the results presented here we are more concerned with the possible variation of $\Delta \lambda_{\mathrm{inst}}$ along the slit than with the absolute value of $\Delta \lambda_{\mathrm{inst}}$, so we use the position-varying values given by \citet{Young:EIS:2011}, which can be accessed through the eis\_slit\_width routine of the EIS analysis software. We have checked the sensitivity of our results to $\Delta \lambda_{\mathrm{inst}}$ by deriving $v_{\mathrm{eff}}$ assuming the smaller, fixed instrumental widths of 0.061~\AA\ and 0.062~\AA\ discussed above and found that the trends in $v_{\mathrm{eff}}$ with height remain essentially the same.

\subsection{Scattered Light: Level}\label{subsec:scatlev}

Superimposed on the true coronal emission is instrumental scattered light, which 
consists of an unshifted ghost spectrum due to scattering of disk radiation by
the microroughness of the instrumental optics. In our data line widths on the disk were
narrower than in the off-disk spectra. Thus, scattered light contamination is expected to reduce
the measured line widths. Far enough above the limb, the instrumental scattered 
light may dominate the measured intensity and cause the observed line width to decrease. 

There are two issues that must be addressed in order to understand the role of stray light
in the analysis. First, we must know the actual level of scattered light in EIS. 
Second, we need to determine how the stray light changes the inferred line widths.

	The scattered light contribution to the EIS spectra was tested during an eclipse where the moon occulted a portion of the solar disk \citep{Ugarte:EIS:2010}. It was found that the stray light intensity in the eclipsed portion of the observation was about 2\% of the intensity in the uneclipsed portion. Previous measurements of the off-disk intensity of lines formed at temperatures well below typical coronal temperatures have supported this value for the stray light level \citep{Hahn:ApJ:2011}. 
	
	In principle, the amount of stray light in the EIS observations depends on the pointing of the instrument as well as on the emission of the solar disk, so that the 2\% level found by \citet{Ugarte:EIS:2010} may not be 
representative of the real level of the scattered light in our data. To estimate the stray light in our
observations we have measured the intensity of a chromospheric line. Outside the solar disk, chromospheric lines
usually do not emit any radiation and the measured intensity is entirely due to scattered light. 

\begin{figure}
\centering \includegraphics[width=0.9\textwidth]{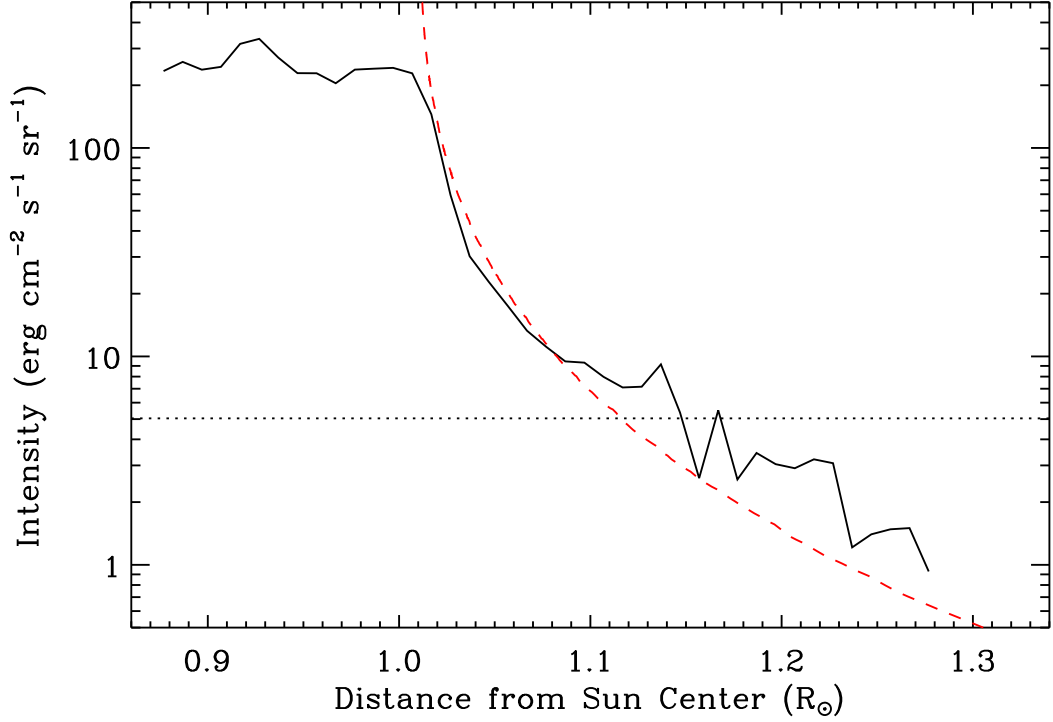}
\caption{\label{fig:he2} The solid line shows the measured He~\textsc{ii}~256.32~\AA\ line intensity in the observation. The horizontal dotted line indicates indicates the intensity that is 2\% of the on-disk average. Above about 1.15~$R_{\sun}$ the measured intensity is clearly less than this value, which is therefore an upper bound on the scattered light. The dashed line illustrates the model prediction for the real emission from the He~\textsc{ii} line. We expect the difference between the measurement and the model to be caused by instrumental scattered light. The close agreement between the measured and predicted emission implies that the scattered light level is below $2\%$ for all of the off-limb data. 
}
\end{figure}

	The coldest line observed by EIS that can be used for this estimate is He~\textsc{ii} 256.23~\AA. This is the only chromospheric line in the EIS spectral range that has enough intensity to be observed to large heights off-disk. It is part of a blend with Si~\textsc{x}~256.38~\AA\ and several weaker coronal lines, but the blended lines all lie on the long wavelength side of the He~\textsc{ii} line and can be easily separated using a multi-Gaussian fit. The solid line in Figure~\ref{fig:he2} shows the He~\textsc{ii} intensity measured in our observation. It is clear that above $1.15$~$R_{\sun}$ the He~\textsc{ii} intensity is less than 2\% of its intensity on-disk. Therefore above $1.15$~$R_{\sun}$ the stray light intensity is $<2\%$ of the on-disk intensity. 
	
	Below 1.15~$R_{\sun}$ and approaching the solar limb, we observe an increasing level of He~\textsc{ii} intensity. However, this turns out to be real emission, not scattered light. This is because of the large elemental abundance of Helium, which allows there to be a significant He~\textsc{ii} abundance in the corona even though the formation temperature of He~\textsc{ii} is well below the coronal temperature. To estimate the expected coronal intensity of the He~\textsc{ii} line we have calculated the ion charge state distribution of Helium in an accelerating fast solar wind using the model of \citet{Cranmer:ApJS:2007}. This model provides the plasma velocity, temperature, and density along the entire trajectory of the solar wind starting from the lower chromosphere. We have used these values to solve the time-dependent equation for the charge state distribution following \citet{Landi:ApJ:2012a}. We then combined this model with the CHIANTI database to calculate the line emissivity at every point along the trajectory. Both collisional and radiative excitation were included in the model. For the resonant radiative excitation we have used the He~\textsc{ii} 256.32~\AA\ line intensity we measured on the solar disk. 
	
	The total line intensity was obtained by integrating the emissivity along the line-of-sight assuming a radially expanding coronal hole, and neglecting the presence of streamers. We also applied a correction for the difference in density between the \citet{Cranmer:ApJS:2007} model and recent density measurements from \citet{Hahn:ApJ:2010}. This correction reduced the predicted line intensity by a factor of $\approx1.5$. 
	
	The dashed line in Figure~\ref{fig:he2} indicates the He~\textsc{ii} 256.32~\AA\ intensity predicted by the model. Below 1.15~$R_{\sun}$ the observed emission is almost identical to the predicted intensity, both in terms of the absolute intensity and the rate of decrease. This agreement is remarkably good, especially considering that the \citet{Cranmer:ApJS:2007} model was developed completely independently from our observations and we have used a simplified coronal geometry. This excellent agreement implies that even below 1.15~$R_{\sun}$ the observed He~\textsc{ii} line intensity is not due to scattered light, but rather by real emission; and that the instrumental scattered light is below the $2\%$ threshold estimated by \citet{Ugarte:EIS:2010}. Based on these results, we conservatively take the stray light level to be $2\%$ of the average on disk intensity for the rest of the analysis. We emphasize that this is an upper limit and the actual stray light level is likely to be even less. 
	
\subsection{Scattered Light: Effect}\label{subsec:scateffect}
	
	To directly account for the stray light when deriving the line widths, we used a sum of two Gaussians to fit the data and extract the line widths. One of the Gaussians in the fit had fixed parameters and represented the scattered light profile while the other had free parameters and represented the real emission. This procedure is equivalent to subtracting a stray light profile from the data. 
	
\begin{figure}
\includegraphics[width=0.9\textwidth]{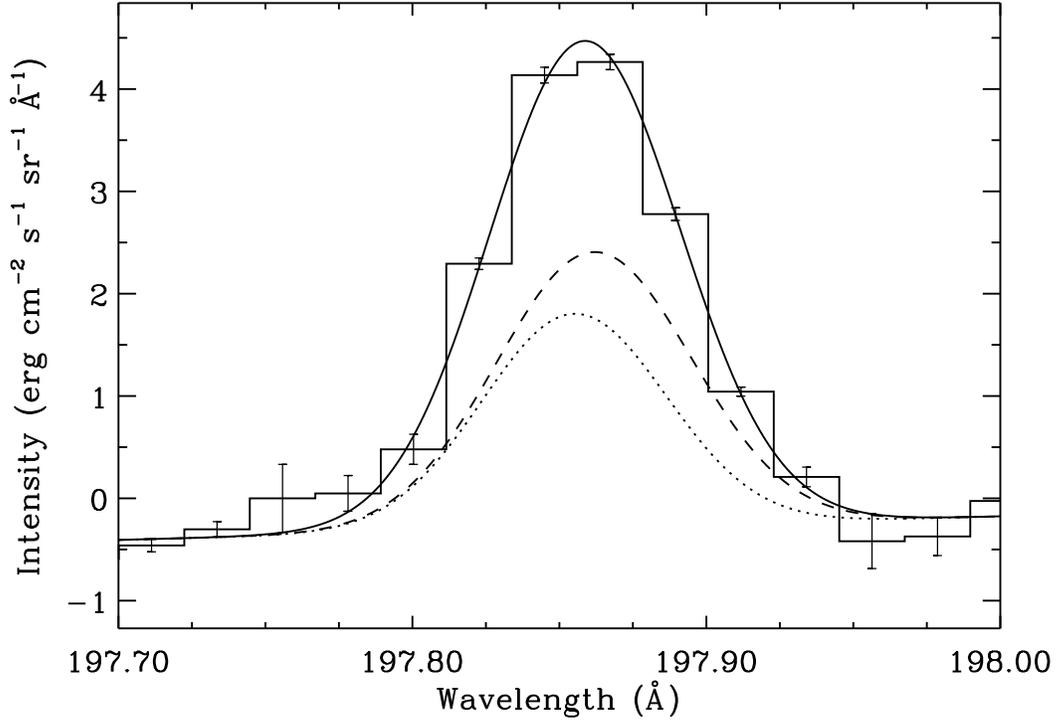}
\caption{\label{fig:fe9profile} Two-Gaussian fit to the Fe~\textsc{ix} 197.86~\AA\ line at $1.3$~$R_{\sun}$. Here the dotted line is the fixed stray light profile with a FWHM $0.074$~\AA, the dashed line is the inferred profile from real emission giving FWHM $0.0803 \pm 0.0068$~\AA, and the solid line is the sum of the two. A single Gaussian fit to the same data would give $\Delta \lambda =0.0772 \pm 0.0024$~\AA. This is an extreme example with a very high stray light level of $42\%$.
}
\end{figure}

	First, a single Gaussian fit was performed to the data below the solar limb. The stray light line intensity above the limb was set to $2\%$ of the on-disk intensity. The line width used for the stray light profile was the same as that extracted from the below limb data, but corrected for the varying instrumental width. The centroid position for the stray light was the same as that measured on the disk. A two-Gaussian fit to the data above the limb was then performed to derive the line width from the real emission. An example of a fit for a case with a very high fraction of scattered light is shown in Figure~\ref{fig:fe9profile}, which illustrates this procedure for the Fe~\textsc{ix}~197.86~\AA\ line at about 1.3~$R_{\sun}$.
	
	In order to estimate the possible influence of scattered light on the inferred line widths we also performed measurements using single Gaussian fits to determine the line widths and compared them to the two-Gaussian fits described above. The widths from the two-Gaussian fits were slightly greater than those from the single Gaussian fits. This is the expected result since we essentially subtracted the narrower stray light profiles from the data. However, the results from either single or double Gaussian fits agreed to within the uncertainties as long as the stray light fraction was less than $\sim 45\%$ of the total intensity. For example, for Fe~\textsc{ix} at 1.3~$R_{\sun}$, where the scattered light fraction is $42\%$, the FWHM of the Fe~\textsc{ix}~197.86~\AA\ line was $\Delta \lambda = 0.0772 \pm 0.0024$~\AA\ from a single Gaussian fit and $\Delta \lambda = 0.0803 \pm 0.0068$~\AA\ from the two-Gaussian fit that corrects for stray light (Figure~\ref{fig:fe9profile}). 

	To see the effect of scattered light in more detail we performed single Gaussian fits on model spectra. A synthetic spectral line profile was generated as the sum of two Gaussians, representing the real and stray light emission. The line widths for the two-Gaussian components were set using typical values measured in off-disk and on-disk portions of the observation. We also considered small shifts in the centroid position of the scattered light component. We then varied the relative intensity of the two components and fit the resulting line using a single Gaussian. We found that the output line width varies smoothly between the ``real'' and ``stray light'' input line widths as the scattered light fraction increases. However, in order to produce decreases in the inferred line width of the magnitude found in our observations, the scattered light must be $\gtrsim 45\%$ of the total intensity. This confirms the result we found from comparing single and double Gaussian fits of the actual data. 
	
\begin{figure}
\centering \includegraphics[width=0.9\textwidth]{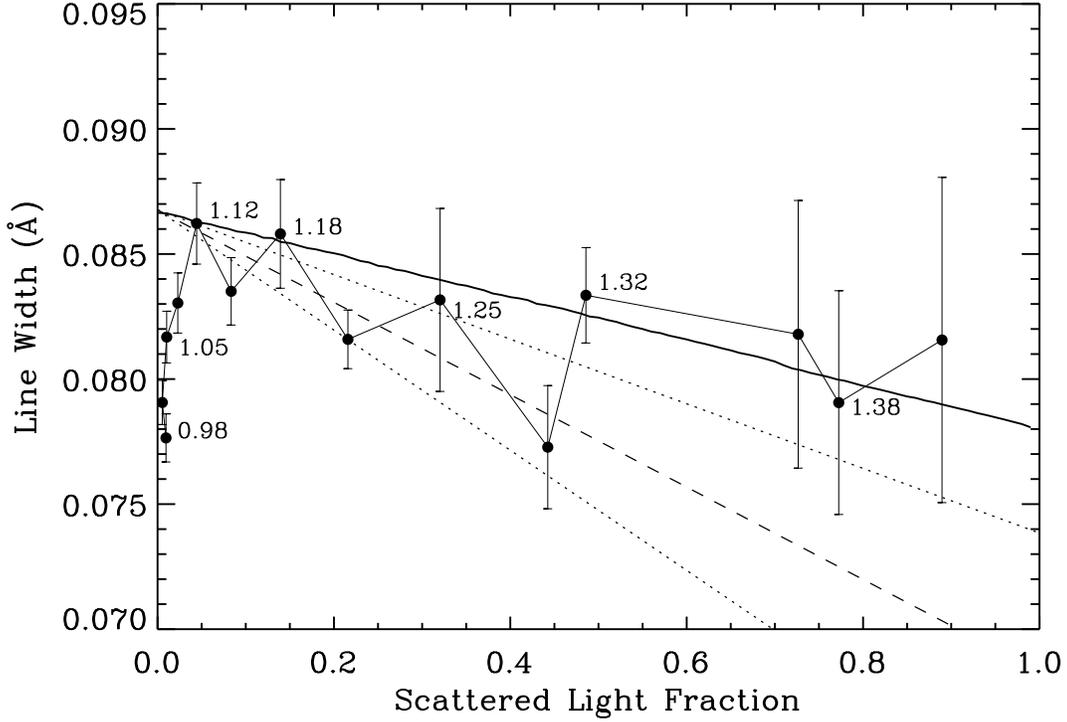}
\caption{\label{fig:Fe9Scat} The thick solid line shows the Gaussian line width that would be inferred from a single Gaussian fit as a function of the scattered light fraction. In this model the line width of the real component is set to the width of the Fe~\textsc{ix} 197.86~\AA\ line at its maximum at 1.12~$R_{\sun}$. The line width of the scattered component is set to the width of the same line on the solar disk. The points on the plot show the observed line width of the Fe~\textsc{ix} line with labels on selected points to indicate the corresponding height in units of $R_{\sun}$. The dashed line is a linear fit to the observations between 1.1~$R_{\sun}$ and 1.3~$R_{\sun}$ to illustrate that the lines narrow more rapidly than predicted by the model when scattered light is less than $\sim 45\%$ of the total line intensity. The $1\sigma$ uncertainty in the slope of the fit is indicated by the dotted lines.}
\end{figure}

	Figure~\ref{fig:Fe9Scat} shows an example of this model for Fe~\textsc{ix} 197.86~\AA, the line with the largest scattered light fraction in the observations. In this example the ``real'' emission line width was set to the maximum observed width, corresponding to 1.12~$R_{\sun}$. This overestimates the influence of scattered light since the real line width should be increasing with height. Even so, the observed line widths decrease more rapidly between $1.1$~$R_{\sun}$ and $1.3$~$R_{\sun}$ than can be explained by scattered light.

	The line widths described throughout the rest of the paper were determined using the two-Gaussian fit to correct for the stray light. In order to ensure that the results we obtain are robust to stray light contamination, we adopt a scattered light fraction cutoff of 45\% and consider only data below this limit. 
			
\section{Results}\label{sec:res}

\begin{figure}
\includegraphics[width=0.9\textwidth]{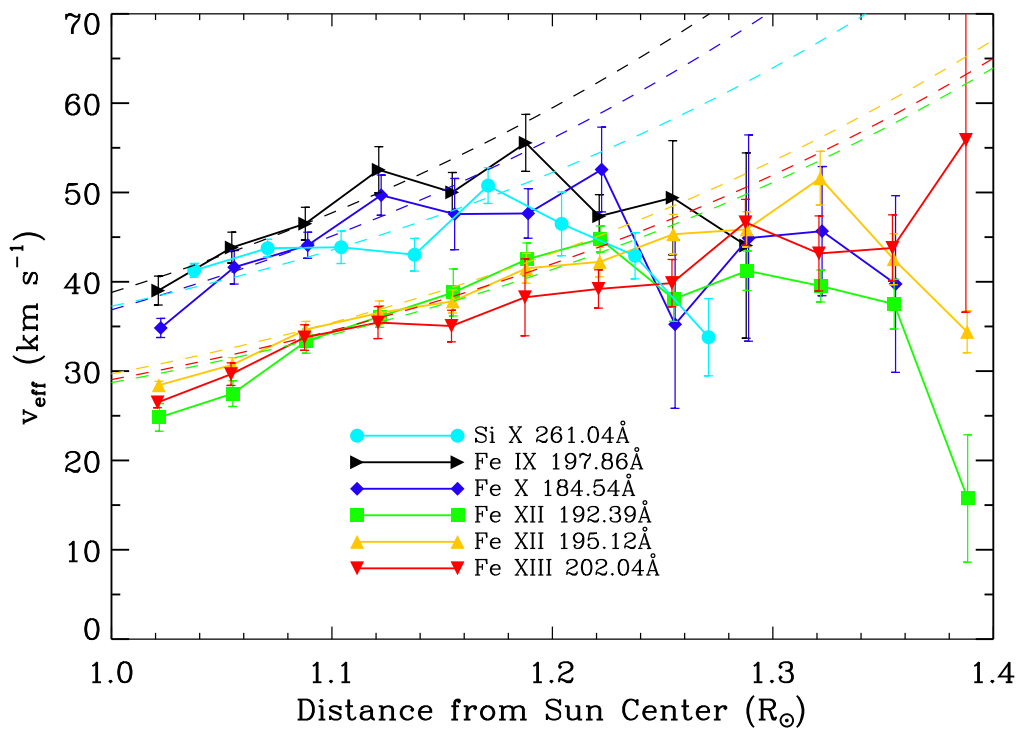}
\caption{\label{fig:VeffR} Line widths, expressed as $v_{\mathrm{eff}}$ versus height. Points where the instrumental scattered light is greater than 45\% have been omitted.  The dashed lines illustrate, for each line, the $v_{\mathrm{eff}}$ expected for undamped waves normalized at 1.1~$R_{\sun}$.}
\end{figure}

Effective velocities were derived from the line widths by subtracting the spatially varying instrumental width as described in Section~\ref{subsec:wid}. Figure~\ref{fig:VeffR} shows $v_{\mathrm{eff}}$ for various spectral lines. We focus on these six 
particular lines because they are strong, relatively isolated in the spectrum, and could be measured reliably to large heights. The apparent offsets in $v_{\mathrm{eff}}$ we attribute to differences in $T_{\mathrm{i}}$, which is a function of the charge-to-mass ratio of the ion \citep{Landi:ApJ:2009,Hahn:ApJ:2010}. The figure shows that $v_{\mathrm{eff}}$ increases initially for all ions, but flattens out and begins to decline between 1.1 -- 1.2~$R_{\sun}$ for Si~\textsc{x}, Fe~\textsc{ix} and Fe~\textsc{x}. The Fe~\textsc{xii} lines also show some evidence for a decrease, but beginning above 1.2~$R_{\sun}$. A clear flattening out is observed for the Fe~\textsc{xiii} line, though no decrease is seen in $v_{\mathrm{eff}}$. 

	If the non-thermal velocity is due to Alfv\'en waves then it is related to the wave amplitude by
$\avg{\delta v^{2}} = 2v_{\mathrm{nt}}^2$ \citep{Hassler:ApJ:1990, Banerjee:AA:1998}. 
The energy flux density of the waves is then given by 
$F=\rho^{1/2} v_{\mathrm{nt}}^2 B/\sqrt{\pi}$, where $\rho$ is the mass density and 
$B$ is the magnetic field strength \citep{Hollweg:CompRep:1990}. \citet{Moran:AA:2001} has shown that
for undamped waves in a flux tube geometry conservation of energy implies that 
$v_{\mathrm{nt}} \propto \rho^{-1/4}$. Thus, in a polar coronal 
hole $v_{\mathrm{nt}}$ should increase with $R$ due to the exponential decrease of 
$\rho$ with height. A constant or decreasing $v_{\mathrm{nt}}$ versus $R$ implies 
wave damping.

The dashed lines in Figure~\ref{fig:VeffR} roughly illustrate the expected $v_{\mathrm{eff}}(R)$ for undamped waves based on the $\rho^{-1/4}$ dependence of $v_{\mathrm{nt}}$, where we have estimated the density from an Fe~\textsc{ix} intensity ratio and assumed $T_{\mathrm{i}}=T_{\mathrm{e}}$ \citep{Hahn:ApJ:2010}. The deviation from these trends suggests that all lines do indeed show indications of damping, including the Fe~\textsc{xii} and Fe~\textsc{xiii} lines even though their line widths do not decrease as strongly with height.

Similar results found in the past were mainly ascribed to two
effects: instrument-scattered light and resonant photoexcitation from radiation emitted 
by the disk. Here we show that neither of these are an issue with our results. We also consider and rule out effects due to the line-of-sight integration implicit in the observations.

\subsection{Scattered Light}\label{subsec:scat}

\begin{figure}
\includegraphics[width=0.9\textwidth]{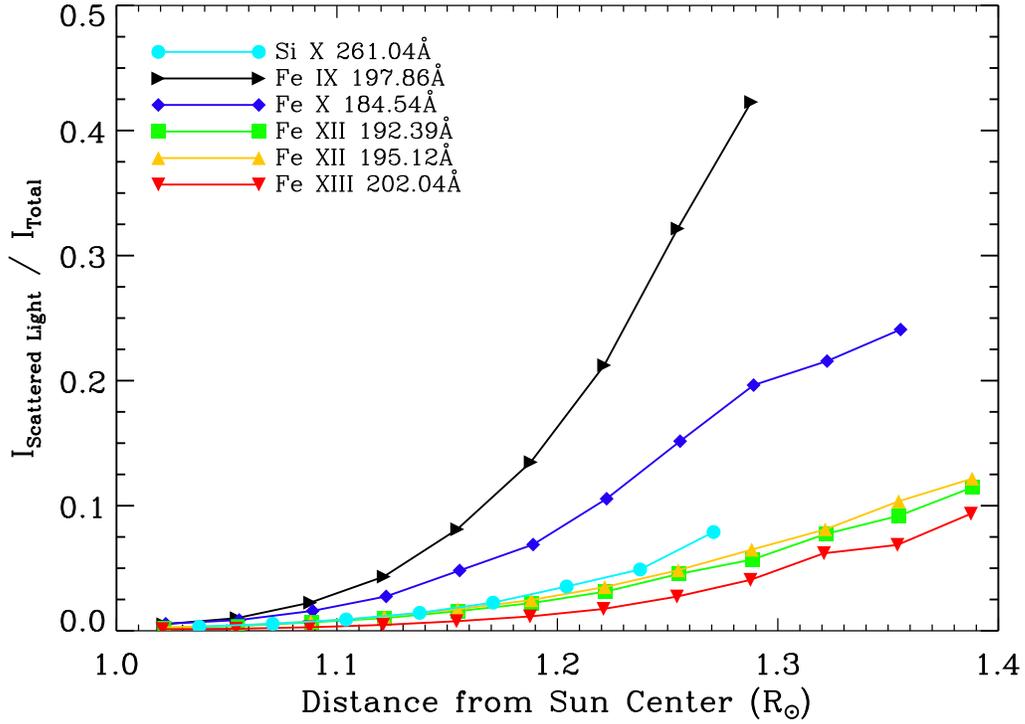}
\caption{\label{fig:Scatlight} Scattered light intensity in each line as a fraction of the total intensity for the line widths shown in Figure~\ref{fig:VeffR}. The scattered light fraction for each line is below about 10\% at the heights where damping begins to be seen in Figure~\ref{fig:VeffR}.}
\end{figure}

	As described in detail in Sections~\ref{subsec:scatlev} and \ref{subsec:scateffect}, we accounted for 
instrument scattered light in our observations by performing two-Guassian fits to the data that included a stray 
light profile based on below limb data. We note that when
stray light contributes $\lesssim 45\%$ of the total intensity, even the line widths inferred from
uncorrected single Gaussian fits were found to give the same results. Figure~\ref{fig:Scatlight} shows the ratio of scattered light to total intensity for each of the lines from Figure~\ref{fig:VeffR}. For each line, stray light makes up less than 10\% of the total intensity at the heights where the downturn in $v_{\mathrm{eff}}$ begins. Thus, even if there were a large uncertainty in the stray light intensity, the scattered light fraction would be too small to explain the observed behavior in $v_{\mathrm{eff}}$ versus $R$.	

\subsection{Photoexcitation}\label{subsec:phot}
	\citet{Oshea:AA:2005} showed that the widths of two Mg~\textsc{x} lines began to decrease at about the same point where photoexcitation became important. They suggested that their results were due to resonant photoexcitation systematically affecting the line width measurements, although the exact mechanism was unspecified. However, we can rule out photoexcitation as an influence on the lines observed here.
	
	We have measured the relative intensity of the three strong Fe~\textsc{xii} lines at 192.39~\AA, 193.51~\AA, and 195.12~\AA. These are sensitive to resonant scattering of disk radiation in a way similar to the commonly used O~\textsc{vi} 1031~\AA/1037~\AA\ intensity ratio \citep{Kohl:AARv:2006}. These Fe~\textsc{xii} lines are due to $3s^2 3p^3\,^{4}S_{3/2} \leftarrow 3s^2 3p^2 (^{3}P) 3d\,^{4}P_{J}$ transitions for $J =$ 1/2, 3/2, 5/2, respectively. The Fe~\textsc{xii}~193.51~\AA\ line blends with an Fe~\textsc{xi} line, which prevents it from being used in the line width analysis. However, the individual intensities of the two blended lines could be determined by estimating the intensity of the Fe~\textsc{xi} blend component using the measured intensity of the Fe~\textsc{xi} 188.30~\AA\ line. Both Fe~\textsc{xi} lines come from the same upper level so their relative intensities are determined solely by a branching ratio.
	
\begin{figure}
\includegraphics[width=0.9\textwidth]{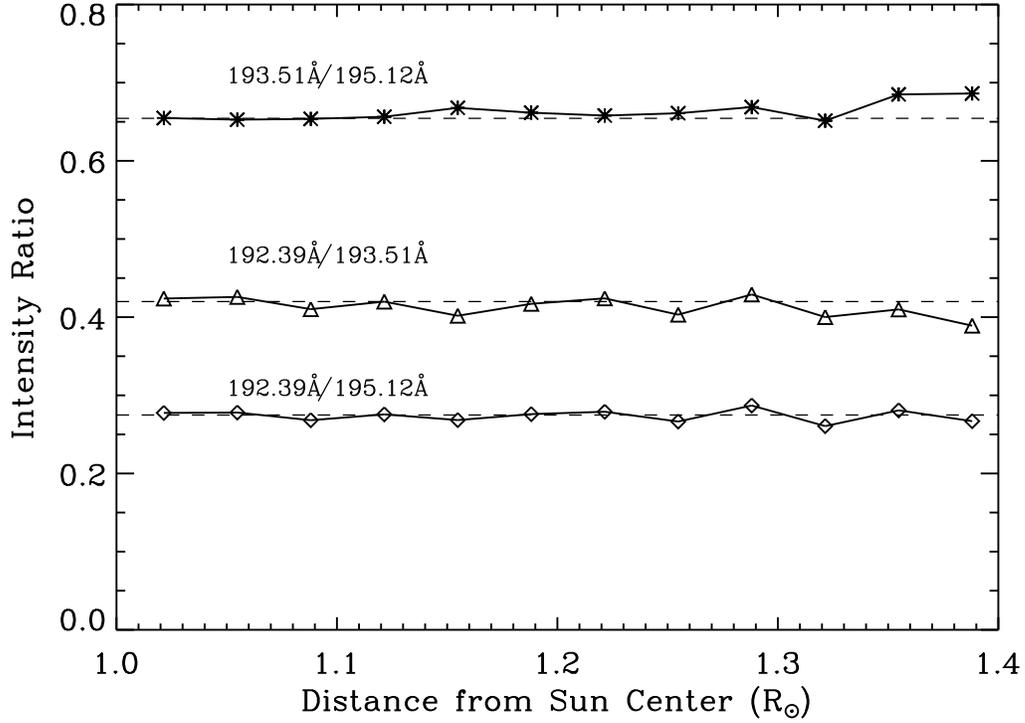}
\caption{\label{fig:Fe12Ratios} The intensity ratios of Fe~\textsc{xii}~$3s^2 3p^3\,^{4}S_{3/2} \leftarrow 3s^2 3p^2 (^{3}P) 3d\,^{4}P_{J}$ for $J = $1/2, 3/2, and 5/2, corresponding to the 192.39~\AA, 193.51~\AA, and 195.12~\AA\ lines, respectively. These line ratios are sensitive to resonant photoexcitation. The dashed lines show the average ratio between 1.05 and 1.15~$R_{\sun}$. If photoexcitation were significant these ratios would decrease with height as the collisional excitation rate drops \citep{Kohl:ApJ:1982}. That the ratios are constant with height demonstrates that photoexcitation does not affect these Fe~\textsc{xii} lines. }
\end{figure}

	Figure~\ref{fig:Fe12Ratios} shows that the ratios among these Fe~\textsc{xii} lines are constant with height. If photoexcitation were important the ratios would decrease with height as the plasma changes from being collisionally to radiatively excited \citep{Kohl:ApJ:1982}. Thus, photoexcitation does not affect the Fe~\textsc{xii} lines. 
	
	The other lines in our study do not have such convenient diagnostics for detecting photoexcitation, but we can infer that it probably is not a factor for these lines either. First, like the Fe~\textsc{xii} lines, the Fe~\textsc{x} and~\textsc{xiii} lines start in upper levels connected to the ground state by a dipole transition. Compared to the Fe~\textsc{xii} lines, the product of the oscillator strengths \citep{Landi:ApJ:2012} and the on-disk intensities is about a factor of three less for the Fe~\textsc{xiii} line and about a factor of three greater for the Fe~\textsc{x} line. Since these are within a factor of a few, the sensitivities of the Fe~\textsc{x} and \textsc{xiii} lines are probably similar to those of the Fe~\textsc{xii} lines and we expect that photoexcitation is unimportant. A similar argument can be made for the Si~\textsc{x} lines used. 
	
	An even stronger argument can be made against photoexcitation influencing the Fe~\textsc{ix} 197.86~\AA\ line. This line is due to  a $3s^23p^53d\,^{1}P_{1} \leftarrow 3s^23p^54p\,^{1}S_{0}$ transition. The ground state of Fe~\textsc{ix} is $3s^23p^6\,^{1}S_{0}$. Thus, photoexcitation from the ground state to this upper level would involve a strictly forbidden $J=0$ -- $0$ radiative transition. Also, the lower level, $3s^23p^53d\,^{1}P_{1}$, is connected to the ground level by a strong dipole transition, so that its population is very low at any density and radiative pumping from this level to the upper $^{1}S_{0}$ level is negligible. Since the $3s^23p^54p\,^{1}S_{0}$ level can essentially only be populated by collisions, photoexcitation is unimportant for this Fe~\textsc{ix} line. 

\subsection{Line-of-Sight}\label{subsec:los}

	Another factor that could influence the observed $v_{\mathrm{eff}}$ is the orientation of the magnetic field along the line-of-sight of the observations. There are two possible effects. First, the outflowing solar wind could have a velocity component along the line-of-sight which would produce broadening. Second, the field lines are not always perpendicular to the line of sight. Hence fluctuations perpendicular to the magnetic field will be tilted relative to the line-of-sight so that the observed $v_{\mathrm{eff}}$ appears smaller. Both possibilities affect the observations in opposite ways. Because the line intensity is proportional to $n_{\mathrm{e}}^2$, which falls rapidly with height, the data are dominated by the nominal observation height, which is the point closest to the Sun and where the line-of-sight is essentially perpendicular to the magnetic field. To estimate the magnitude of these effects we assumed that the magnetic field lines in the polar coronal hole are radial, that the density scale height is $\sim 0.07$~$R_{\sun}$ (corresponding to a typical coronal hole temperature of $\log T(\mathrm{K}) = 6.0$) and that the outflow velocity is $\approx 13$~$\mathrm{km\,s^{-1}}$ \citep{Cranmer:ApJ:1999}. With these values we estimate that each of these effects may cause line broadening or narrowing on the order of $\Delta v_{\mathrm{eff}} \sim 0.5$~$\mathrm{km\,s^{-1}}$. This is about 1\% the size of the measured values for $v_{\mathrm{eff}}$ and thus unimportant. To some extent the effects can also be expected to cancel each other out.
		
\subsection{Differential Emission Measure}\label{subsec:dem}

	There are some indications that the observation may include multiple plasma structures. The profiles for $v_{\mathrm{eff}}$ versus $R$ in Figure~\ref{fig:VeffR} display a dependence on the ion formation temperature. The Fe~\textsc{ix}, Fe~\textsc{x} and Si~\textsc{x} lines form at cooler temperatures than Fe~\textsc{xii} or Fe~\textsc{xiii} \citep{Bryans:ApJ:2009} and show an earlier and more pronounced drop in $v_{\mathrm{eff}}$ with height. To study this in more detail we have performed a Differential Emission Measure (DEM) analysis using the technique described in \citet{Landi:AA:1997} and \citet{Hahn:ApJ:2011}. The DEM, $\phi(T)$, shows the distribution of material along the line-of-sight as a function of the electron temperature $T$. 

\begin{figure}
\includegraphics[width=0.9\textwidth]{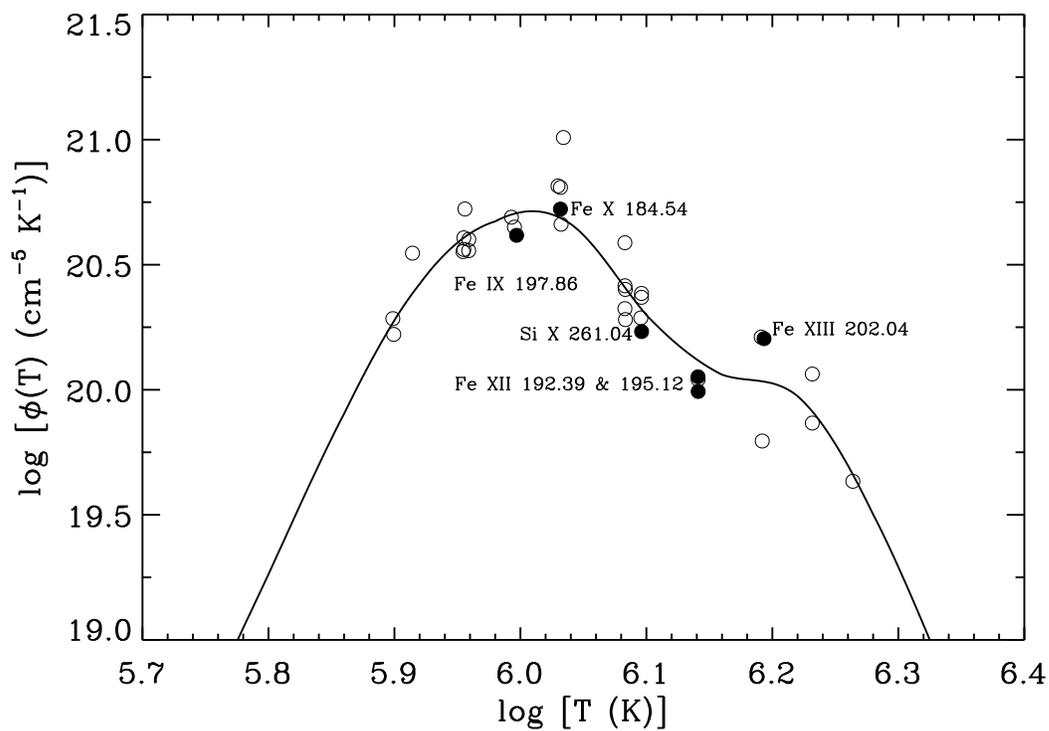}
\caption{\label{fig:DEM} The DEM at 1.1~$R_{\sun}$. The open and filled circles show the points based on the measured line intensities used to calculate the DEM. The scatter in these points gives an estimate of the uncertainty in the DEM. The filled circles highlight those particular points corresponding to the lines in Figure~\ref{fig:VeffR}. }
\end{figure}

Figure~\ref{fig:DEM} shows $\phi(T)$ at 1.1~$R_{\sun}$. The circles on the plot indicate the points 
$\phi(T_{t})$ determined by the measured line intensities, where $T_{t}$ essentially 
represents the average temperature of the plasma from which each emission line originates. 
The filled circles highlight the points corresponding to the lines used to 
determine $v_{\mathrm{eff}}(R)$. The DEM shows that the coronal hole emission is dominated 
by plasma around $\log T(\mathrm{K}) = 6.0$, but there is a high temperature tail at 
$\log T(\mathrm{K}) = 6.1$ -- $6.2$. This form of DEM implies that there are multiple 
structures along the line-of-sight. Among the possible interpretations are that the 
cool peak represents emission from the polar coronal hole while the warm tail comes from surrounding 
streamer plasma. Alternatively, the DEM may indicate that there are 
distinct structures with different temperatures within the polar coronal hole. 

The shape of the DEM may explain the observed differences in the $v_{\mathrm{eff}}(R)$ 
behavior of the different ions. The fraction of emission originating in the different parts
of the DEM can be quantified by integrating over the DEM and the contribution function of 
each line. For the purpose of this estimate we take $\log T(\mathrm{K}) = 6.1$ to be the dividing
line between the cool and the warm structure. Integrating from this point up to higher temperatures
we find that the fraction of emission coming from the high temperature material is 5\%, 14\%, and 43\% 
of the total intensity for 
Fe~\textsc{ix}, Fe~\textsc{x}, and Si~\textsc{x}, respectively and so these lines come
primarily from the low temperature peak of the DEM. For Fe~\textsc{xii} and Fe~\textsc{xiii} 
the fraction from the high temperature tail is 67\% and 90\% of the total intensity, respectively. This 
suggests that the different structures along the line-of-sight could have different 
$v_{\mathrm{eff}}(R)$ profiles as we discuss below.

Line widths in coronal streamers have been observed to be narrower than in coronal holes \citep{Dolla:AA:2008}. 
If the warm material in the DEM represents intervening streamer plasma then 
contamination of the Si~\textsc{x}, Fe~\textsc{ix} and Fe~\textsc{x} by emission from that structure could
systematically influence the inferred line widths. However, there are several factors that 
indicate that these line widths are dominated by the cool material, presumably from the coronal hole. The situation
here is analogous to the systematic effect of instrumental stray light. In that case we showed that
more than $45\%$ of the intensity must come from the stray light before the line widths are 
significantly changed. The fraction of light from the warm material is definitely less than $45\%$ for
Fe~\textsc{ix} and Fe~\textsc{x}. Additionally, the Si~\textsc{x}, Fe~\textsc{ix} and Fe~\textsc{x} 
lines all show the same behavior, whereas if there
were a systematic error from contamination by emission from the warm structure then they would behave
differently according to the varying level of emission. Therefore, we conclude that
the velocities derived from Fe~\textsc{ix}, Fe~\textsc{x}, and Si~\textsc{x} 
reflect the properties of the polar coronal hole. 

\section{Discussion}\label{sec:dis}

Our analysis shows unambiguous evidence that $v_{\mathrm{nt}}$ decreases at relatively low heights in a polar coronal hole. These results avoid or resolve the uncertainties that have affected previous measurements. Since $v_{\mathrm{nt}}$ is thought to be proportional to the Alfv\'en wave amplitude, these observations imply that Alfv\'en waves are indeed damped at low heights in coronal holes. Previous studies with SUMER, CDS, and now EIS have all shown decreases in $v_{\mathrm{eff}}$ at low heights \citep{Banerjee:AA:1998,Doyle:AA:1999,Moran:ApJ:2003,Oshea:AA:2005,Dolla:AA:2008}. Results have been consistent across different instruments with varying stray light characteristics and across ions and particular emission lines with varying sensitivity to resonant photoexcitation. This is further evidence that such proposed systematic issues are unlikely to explain the observed decrease. It seems that previous researchers have overestimated the importance of the systematic uncertainties on their results. The present results are particularly consistent with those of \citet{Moran:ApJ:2003}, where an apparent dependence on formation temperature was also noted. 

Based on the DEM analysis, our results imply that Alfv\'en 
waves are dissipated within the cooler structure and that in the warmer structure the damping is 
less pronounced. If the warmer structures are interpreted as streamer plasma along the line-of-sight,
our results indicate that Alfv\'en wave damping is active at low heights in the coronal
hole and hence in the fast wind, but is not as strong in streamer material where the slow wind is believed to originate. 
If Alfv\'en wave damping occurs via turbulent cascades of wave power from low frequencies to frequencies 
high enough to cause ion-cyclotron acceleration, our results are consistent with signatures of ion-cyclotron effects
being measured in coronal holes \citep{Landi:ApJ:2009} and not in streamers 
\citep{Landi:ApJ:2007}.
	
Though dissipation of Alfv\'en waves by viscosity, thermal conductivity, and resistivity 
are not expected to be important below 2~$R_{\sun}$ \citep{Parker:ApJ:1991, Cranmer:SSR:2002}, 
damping may be enhanced in an inhomogenous plasma. For 
example, it has been shown that phase-mixing induced by adjacent Alfv\'en waves on 
neighboring field lines resonating out of phase with each other can enhance the viscous and 
resistive dissipation rates \citep{Heyvaerts:AA:1983}. Or the interaction of 
outward and inward propagating Alfv\'en waves can 
lead to a turbulent cascade that transfers energy to higher frequencies and forms other 
types of plasma waves, which are more strongly damped \citep{Matthaeus:ApJ:1999}. 

Regardless of the physical mechanism that leads to the wave dissipation, the energy 
must be deposited in the low corona and contribute to coronal heating.
We can estimate an upper bound for the dissipated energy using the expression for the energy flux of an Alfv\'en wave 
$F=2\rho~v_{\mathrm{nt}}^2V_{\mathrm{A}}$ \citep{Doyle:SolPhys:1998}, 
where $V_{\mathrm{A}}=B/\sqrt{4\pi\rho}$ is the Alfv\'en speed. The electron density was measured from an Fe~\textsc{ix} line 
intensity ratio to be $\approx 6\times10^{7}$~cm$^{-3}$ at 1.1~$R_{\sun}$. Due to the low intensity, the ratio 
could not be used directly above $\sim 1.15$~$R_{\sun}$, but extrapolating with a scale-height exponential decay 
gives $n_{\mathrm{e}} \approx 9\times10^{6}$~cm$^{-3}$ at 1.3~$R_{\sun}$. This value agrees with 
measurements for polar coronal holes \citep{Wilhelm:ApJ:1998}; and since $F$ varies as $\sqrt{n_{\mathrm{e}}}$,
the uncertainty incurred by extrapolating the density should not seriously affect our estimate.
To estimate the spatially varying magnetic field strength for a super-radially expanding polar coronal hole we used 
an empirical model \citep{Cranmer:ApJ:1999} and take $B\approx7$~G at 1~$R_{\sun}$ \citep{Wang:ApJ:2010}. 
We find that $V_{\mathrm{A}}(1.1\,R_{\sun}) \approx 1200$~$\mathrm{km\,s^{-1}}$ and 
$V_{\mathrm{A}}(1.3\,R_{\sun}) \approx 1600$~$\mathrm{km\,s^{-1}}$. 
We obtain an upper bound on $v_{\mathrm{nt}}$
using the lower bound on the ion temperature of $T_{\mathrm{i}} \approx T_{\mathrm{e}} \approx 10^{6}$~K 
\citep{Landi:ApJ:2009,Hahn:ApJ:2010}. Using Equation~2 we find from the Fe~\textsc{x} lines that $v_{\mathrm{nt}}=42.6$~$\mathrm{km\,s^{-1}}$ 
and 41.8~$\mathrm{km\,s^{-1}}$ at 1.1~$R_{\sun}$ and 1.3~$R_{\sun}$, respectively. 
Similar results are seen for Fe~\textsc{ix} and Si~\textsc{x}. 
Thus, we obtain upper bounds of $F=5.4\times10^{5}$~erg~cm$^{-2}\,$s$^{-1}$ at 
1.1~$R_{\sun}$ and $F=9.2 \times 10^{4}$~erg~cm$^{-2}\,$s$^{-1}$ at 1.3~$R_{\sun}$. As described in Section~\ref{sec:ana}, there is some uncertainty in the EIS instrumental line width and using the smaller estimate would change
these values to $F=7.0\times10^{5}$~erg~cm$^{-2}\,$s$^{-1}$ and 
$F=1.1\times10^{5}$~erg~cm$^{-2}\,$s$^{-1}$ at 1.1~$R_{\sun}$ and 1.3~$R_{\sun}$, 
respectively. This represents an energy flux loss of $5.9 \times 10^5$~erg~cm$^{-2}\,$s$^{-1}$.
The amount of energy required to heat the coronal hole and accelerate 
the fast solar wind is estimated to be $8\times10^{5}$ erg~cm$^{-2}\,$s$^{-1}$ \citep{Withbroe:ARAA:1977}. 
Thus, we estimate that in our observations Alfv\'en waves may deposit up to about $70$\% 
of the energy required to heat the coronal hole and accelerate the fast solar wind. 
The challenge for the field is now to derive an observational lower limit for the energy deposited by Alfv\'en 
waves in coronal holes. Only then will we be able to determine the fraction of heating due to Alfv\'en waves and that
due to magnetic reconnection. 

\section{Summary} \label{sec:sum}

	We have measured the variation of spectral line widths from 1.05 -- 1.40~$R_{\sun}$ over a polar coronal hole. These line widths are thought to be proportional to the Alfv\'en wave amplitude. We found that they deviate from the predicted $\rho^{-1/4}$ dependence for undamped waves. We have investigated possible systematic effects such as instrumental scattered light, resonant photoexcitation, and line-of-sight observational effects and determined that they are all too small to explain our observations. Thus, our results suggest that Alfv\'en waves are damped at surprisingly low heights in a polar coronal hole. We estimate that the amount of energy dissipated can account for a large fraction of that required to heat the coronal hole and accelerate the solar wind. 	
		
\acknowledgements
We thank the anonymous referee for helpful suggestions. MH and DWS were supported in part by the NASA Solar Heliospheric Physics 
program grant NNX09AB25G and the NSF Division of Atmospheric and Geospace Sciences SHINE program
grant AGS-1060194. The work of EL is supported by the NASA grants 
NNX10AQ58G and NNX11AC20G. 
 	

\begin{center}
\begin{longtable}{llc c rcl}
\caption{Line List for DEM Analysis. \label{table:linelist}}
\\
& Ion & & $\lambda$ (\AA)\tablenotemark{1} & \multicolumn{3}{c}{Transition\tablenotemark{1}} \\ \hline
\endfirsthead
\hline
\endhead
\hline
\endlastfoot
 &Mg \textsc{vi} & &268.991 &  $2s^2\, 2p^3\, ^{2}D_{3/2}$ & $-$ & $2s\, 2p^4\, ^{2}P_{1/2}$ \\
	& & &270.391 & $2s^2\, 2p^3\, ^{2}D_{5/2}$ & $-$ & $2s\, 2p^4\, ^{2}P_{3/2}$ \\
 &\raisebox{2.5ex}[0pt]{Mg \textsc{vi}} &\raisebox{2.5ex}[0pt]{$\Big\{$} &270.400 & $2s^2\, 2p^3\, ^{2}D_{3/2}$ & $-$ & $2s\, 2p^4\, ^{2}P_{3/2}$ \\ 
 &Mg \textsc{vii} & &276.154 & $2s^2\, 2p^2\, ^{3}P_{0}$ & $-$ & $2s\, 2p^3\, ^{3}S_{1}$ \\
 &Si \textsc{vi} & &249.125 & $2s2\, 2p5\, ^{2}P_{1/2}$ & $-$ & $2s\,2p6\,^{2}S_{1/2}$ \\
 &Si \textsc{vii} & &272.648 & $ 2s^2\, 2p^4\, ^{3}P_2 $ &$-$ & $2s\, 2p^5\, ^{3}P_1 $ \\
 &Si \textsc{vii} & &275.361 & $ 2s^2\, 2p^4\, ^{3}P_2 $ &$-$ & $2s\, 2p^5\, ^{3}P_2 $ \\
 &Si \textsc{vii} & &275.676 & $ 2s^2\, 2p^5\, ^{3}P_1 $ &$-$ & $2s\, 2p^5\, ^{3}P_1 $ \\
 &Si \textsc{ix} & &258.082 & $ 2s^2\, 2p^2\, ^{1}D_2 $ &$-$ & $2s\, 2p^3\, ^{1}D_2 $\\
 &Si \textsc{x} & &258.374 & $ 2s^2\, 2p\, ^2P_{3/2} $ &$-$ & $2s\, 2p^2\, ^{2}P_{3/2} $ \\
$\ast$&Si \textsc{x} & &261.057 & $2s^2\, 2p\, ^2P_{3/2} $ &$-$ & $2s\, 2p^2\, ^{2}P_{1/2} $ \\ 
 &Si \textsc{x} & &271.992 & $ 2s^2\, 2p\, ^2P_{1/2} $ &$-$ & $2s\, 2p^2\, ^{2}S_{1/2} $ \\
 &Si \textsc{x} & &277.264 & $ 2s^2\, 2p\, ^{2}P_{3/2} $ &$-$ & $ 2s\, 2p^2\, ^{2}S_{1/2} $ \\
 &Fe \textsc{viii} & &185.213 & $ 3p^6\, 3d\, ^{2}D_{5/2} $ &$-$ & $3p^5\,3d^2\, (^{3}F)\, ^{2}F_{7/2}$ \\
 &Fe \textsc{viii} & & 186.599 & $3p^6 \,3d\, ^{2}D_{3/2} $ &$-$ & $3p^5\, 3d^2\, (^{3}F)\, ^{2}F_{5/2}$ \\
 &Fe \textsc{viii} & & 194.661 & $ 3p^6 \, 3d\, ^{2}D_{5/2} $ &$-$ & $3p^6\, 4p\, ^{2}P_{3/2}$ \\
 &Fe \textsc{ix} & &188.497 & $ 3s^2\, 3p^5\, 3d\, ^{3}F_4 $ &$-$ & $3s^2\, 3p^4\, (^{3}P)\, 3d^2\, ^{3}G_5 $\\
 &Fe \textsc{ix} & &189.941 & $ 3s^2\, 3p^5\, 3d\, ^{3}F_3 $ &$-$ & $3s^2\, 3p^4\, (^{3}P)\, 3d^2\, ^{3}G_4 $\\
 $\ast$&Fe \textsc{ix} & &197.862 & $ 3s^2\, 3p^5\, 3d\, ^{1}P_1 $ &$-$ & $3s^2\, 3p^5\, 4p\, ^{1}S_0 $ \\
 &Fe \textsc{x} & &174.531 & $3s^2\, 3p^5\, ^{2}P_{1/2} $ &$-$ & $3s^2\, 3p^4\, (^{3}P)\, 3d\, ^{2}D_{5/2} $\\
 $\ast$&Fe \textsc{x} & &184.537 & $ 3s^2\, 3p^5\, ^{2}P_{3/2} $ &$-$ & $3s^2\, 3p^4\, 	(^1D)\, 3d\,  ^{2}S_{1/2} $ \\
 &Fe \textsc{x} & &190.037 & $ 3s^2\, 3p^5\, ^{2}P_{1/2} $ &$-$ & $3s^2\, 3p^4\, (^{1}D)\, 3d\, ^{2}S_{1/2} $ \\
 &Fe \textsc{x} & &193.715 & $ 3s^2\, 3p^5\, ^{2}P_{3/2} $ &$-$ & $3s^2\, 3p^4\, (^{1}S)\, 3d\, ^{2}D_{5/2} $ \\
														&	&	& 257.259 & $ 3s^2\, 3p^5\, ^{2}P_{3/2} $ &$-$ & $3s^2\, 3p^4\, (^{3}P)\, 3d\, ^{4}D_{5/2} $ \\*
 &\raisebox{2.5ex}[0pt]{Fe \textsc{x}}&\raisebox{2.5ex}[0pt]{$\Big\{$} & 257.263 & $3s^2\, 3p^5\, ^{2}P_{3/2}$  &$-$ & $3s^2\, 3p^4\, (^{3}P)\, 3d\, ^{4}D_{7/2}$\\
 &Fe \textsc{xi} & &180.401 & $ 3s^2\, 3p^4\, ^{3}P_2 $ &$-$ & $3s^2\, 3p^3\, (^{4}S)\, 3d\, ^{3}D_3 $\\
 &Fe \textsc{xi} & &182.167 & $ 3s^2\, 3p^4\, ^{3}P_1 $ &$-$ & $3s^2\, 3p^3\, (^{4}S)\, 3d\, ^{3}D_2 $ \\
 &Fe \textsc{xi} & &188.217 & $ 3s^2\, 3p^4\, ^{3}P_2 $ &$-$ & $3s^2\, 3p^3\, ( ^{2}D)\, 3d\, ^{3}P_2 $ \\
 &Fe \textsc{xi} & &188.299 & $ 3s^2\, 3p^4\, ^{3}P_2 $ &$-$ & $3s^2\, 3p^3\, (^{2}D)\, 3d\, ^{1}P_1 $ \\
 &Fe \textsc{xi} & &189.711 & $3s^2\, 3p^4\, ^{3}P_{0}$ &$-$ & $3s^2\, 3p^3\, (^{2}D)\, 3d\, ^{3}P_{1}$ \\
$\ast$&Fe \textsc{xii} & &192.394 & $3s^2\, 3p^3\, ^{4}S_{3/2}$ &$-$ & $3s^2\, 3p^2\, (^{3}P)\, 3d\, ^{4}P_{1/2}$ \\
 &Fe \textsc{xii} & &193.509 & $3s^2\, 3p^3\, ^{4}S_{3/2}$ &$-$ & $3s^2\, 3p^2\, (^{3}P)\, 3d\, ^{4}P_{3/2}$ \\
$\ast$&Fe \textsc{xii}	& &195.119 & $ 3s^2\, 3p^3\, ^{4}S_{3/2} $ &$-$ & $3s^2\, 3p^2\, (^{3}P)\, 3d\, ^{4}P_{5/2} $ \\
$\ast$&Fe \textsc{xiii} & &202.044 & $ 3s^2\, 3p^2\, ^{3}P_0 $ &$-$ & $3s^2\, 3p\, 3d\, ^{3}P_1 $ \\
    &  & &203.797 & $ 3s^2\, 3p^2\, ^{3}P_{2}$ &$-$ & $ 3s^2\, 3p\, 3d\, ^{3}D_{2}$ \\*
 &\raisebox{2.5ex}[0pt]{Fe \textsc{xiii}}	&\raisebox{2.5ex}[0pt]{$\Big\{$} &203.827 & $ 3s^2\, 3p^2\, ^{3}P_{2}$ &$-$ & $ 3s^2\, 3p\, 3d\, ^{3}D_{3}$ \\
 &Fe \textsc{xiii}& &251.953 & $3s^2\, 3p^2\, ^{3}P_{2}$ & $-$ & $3s\, 3p^{3}\, ^{3}S_{1}$ \\
 &Fe \textsc{xiv} & &264.789 & $3s^2\, 3p\, ^{2}P_{3/2}$ & $-$ & $3s\, 3p^{2}\, ^{2}P_{3/2}$ \\
 &Fe \textsc{xiv} & &270.521 & $3s^2\, 3p\, ^{2}P_{3/2}$ & $-$ & $3s\, 3p^{2}\, ^{2}P_{1/2}$ \\
 &Fe \textsc{xv} & &284.163 & $3s^2\, ^{1}S_{0}$ & $-$ & $3s\, 3p\, ^{1}P_{1}$ \\
\footnotetext[1]{Wavelengths and transitions taken from CHIANTI \citep{Dere:AA:1997,Landi:ApJ:2012}.}
\footnotetext[0]{Brackets indicate blends from the same ion. \\ 
Asterisks mark the lines used in the analysis of the line widths.}
\end{longtable}
\end{center}

\bibliography{CH_LineWidth_arxiv}

\end{document}